\documentclass{article}
\pdfoutput=1
\usepackage[utf8]{inputenc}
\usepackage[T1]{fontenc}
\usepackage{float}
\usepackage{authblk}
\usepackage[normalem]{ulem}
\usepackage{color, colortbl}
\usepackage{amsmath}
\usepackage{amsfonts}
\usepackage{amsthm}
\usepackage{adjustbox}
\usepackage{graphicx}
\usepackage{mathtools}
\usepackage{fullpage}
\usepackage{todonotes}
\usepackage{subcaption}
\usepackage[mathlines]{lineno}
\usepackage[style=numeric,sorting=none]{biblatex}
\definecolor{underbrace}{rgb}{0.04, 0.49, 0.65}
\definecolor{gragreblu}{rgb}{0.04, 0.49, 0.65}
\usepackage[colorlinks = true,
            linkcolor = gragreblu,
            urlcolor  = gragreblu,
            citecolor = gragreblu]{hyperref}
\newcolumntype{P}[1]{>{\centering\arraybackslash}p{#1}}

\title{Finding the right scale of a network: Efficient identification of causal emergence in preferential attachment networks through spectral clustering}
\author[1]{Ross Griebenow}
\author[2,3]{Brennan Klein}
\author[4]{Erik Hoel\thanks{erik.hoel@tufts.edu}}
\affil[1]{Drexel University, Philadelphia, PA, USA}
\affil[2]{Network Science Institute, Northeastern University, Boston, MA, USA}
\affil[3]{Laboratory for the Modeling of Biological and Socio-Technical Systems, Northeastern University, Boston, MA, USA}
\affil[4]{Allen Discovery Center, Tufts University, Medford, MA, USA}

\begin{document}
\maketitle
\pagenumbering{arabic}

\begin{abstract}
All networks can be analyzed at multiple scales. A higher scale of a network is made up of macro-nodes: subgraphs that have been grouped into individual nodes. Recasting a network at higher scales can have useful effects, such as decreasing the uncertainty in the movement of random walkers across the network while also decreasing the size of the network. However, the task of finding such a macroscale representation is computationally difficult, as the set of all possible scales of a network grows exponentially with the number of nodes. Here we compare various methods for finding the most informative scale of preferential attachment networks, discovering that an approach based on spectral analysis outperforms greedy and gradient descent-based methods. We then use this procedure to show how several structural properties of  these networks vary across scales. We describe how meso- and macroscale representations of networks can have significant benefits over their underlying microscale in terms of information transmission, which include properties such as increase in determinism, a decrease in degeneracy, a lower entropy rate of random walkers on the network, an increase in global network efficiency, and higher values for a variety of centrality measures than the microscale.
\end{abstract}

\section{\label{sec:introduction}Introduction}

Networks can be used to represent a wide range of systems, and over the past decade their use has become more common throughout the sciences \cite{barabasi2016network}.  While network analysis is typically performed on the full, microscale representation of a network, our recent research has shown that informative higher scales of networks can be identified and explicitly modeled \cite{klein2020uncertainty}. Using these techniques, a network, $G$, can be recast into a new network, $G_M$, wherein subgraphs of the networks are grouped into individual \textit{macro-nodes}. These macro-nodes summarize the behavior of the subgraph in a manner that recapitulates the dynamics of the original networks. Thus, higher scales act like consistent but dimensionally-reduced models of the original system.

There has long been the assumption in science that, whenever possible, coarse-grained models should be replaced by fine-grained models \cite{gallagher1999beyond}. Due to the general success of this reductionist method, there has been little attention to the gains that accompany macroscale models. This is exacerbated by the lack of formal methods for dealing with systems across scales, as well as the computational cost of detecting an informative higher scale---the number of possible dimension reductions exponentially increases with the size of the system. So while explicitly modeling higher scales has been investigated by the authors in logic gates \cite{hoel2013quantifying, hoel2016can} and Markov processes \cite{hoel2017map}, these investigations have been limited by computational resources.

Here we compare and evaluate different methods for finding informative higher scales of networks by searching the computationally challenging space of possible scales. We compare a greedy algorithm, an approach based on gradient descent, and an approach based on the spectral decomposition of a network. We use these algorithms to show that mesoscale structures are, in general, the most computationally difficult scales to identify in a network. However, this issue is mostly avoided by  the adapted spectral approach introduced here, which has the strongest performance of the algorithms, and can find informative and complex higher scales even in large networks. This spectral analysis technique groups micro-nodes into macro-nodes based on a clustering and ordering approach commonly used in machine learning tasks \cite{OPTICS}. We also explore how higher scales can possess more informative connectivity (a phenomenon called \textit{causal emergence} \cite{hoel2013quantifying}). Since a macroscale is a recasting of an original network, it is critically important to understand how the macroscale networks that are identified by our analysis are different in their network properties compared to their original networks. We find that, compared to their underlying microscale, the macroscales that are found have significant changes to their properties. These include, among others, the decrease the uncertainty of random walkers, an increase the global efficiency of the network, and an increase in various centrality measures of nodes.

\section{\label{sec:methods}Methods}

\subsection{\label{sec:effectiveinformation}Causal emergence in networks}

Every node $v_i$, in a directed network of size $N$ is associated with an out-weight vector, $W^{out}_{i}$, which represents the possible outputs from $v_i$. $W^{out}_{i}$ consists of weights, $w_{ij}$, between node $v_i$ and its neighbors, $v_j$, such that if $w_{ij}=0.0$ there is no edge from $v_i$ to $v_j$. For these networks the $W^{out}_{i}$ of each node sums to $1.0$. Therefore, the edge weights $w_{ij}$ are equivalent to the probability that a random walker on $v_i$ will transition to $v_j$ in the next time step.

In order to find the maximally informative scale of the network, as in \cite{klein2020uncertainty}, we make use of the \textit{effective information} ($EI$), which is a network measure that quantifies the degree of certainty in the transitions of random walkers between nodes in a network, as well as how that certainty is distributed. Therefore, $EI$ is based on two uncertainties. The first is the Shannon entropy of the average out-weight vector in the network, $\langle W_i^{out} \rangle$, which captures how distributed out-weights of the network are. The second is the average entropy of \textit{each} node's $W^{out}_{i}$. Their difference is the $EI$ of a network, such that $EI = H(\langle W^{out}_{i} \rangle) - \langle {H}(W^{out}_{i}) \rangle$. Expanded, this is:

\begin{equation}\label{eq:ei}
    EI = -\sum_{i=1}^N \left( \left(\frac{1}{N}\sum_{j=1}^N w_{ij}{\log_2}\right) 
    \left({\frac{1}{N}\sum_{j=1}^n w_{ij}}\right) \right)-
    \frac{1}{N}\sum_{i=1}^N \left( -\sum_{j=1}^N w_{ij} \log_2 w_{ij}\right)
\end{equation}

When a network has a macroscale with greater $EI$ than its underlying microscale, this is known as \textit{causal emergence}. Causal emergence is when some recast network, $G_M$ (the macroscale), is associated with a gain in $EI$ relative to the original network, $G$ (the microscale). The amount of causal emergence is the difference between the $EI$ of the microscale and that of the macroscale.

\begin{equation}\label{eq:ei}
   \textrm{\textit{causal emergence}} = EI(G_M) - EI(G):
\end{equation}

Note that causal emergence can be negative, in that no possible consistent macro-nodes increase the $EI$, in which case it is referred to as \textit{causal reduction}. Ideally, one would find the macroscale mapping that maximizes the $EI$ of the network, $EI^{max}$, and use the resulting macroscale network to model the system in question. This $EI^{max}$ approximates the channel capacity of the system \cite{hoel2017map}. Note that all networks considered herein are directed. In the case where a network is undirected, $EI$ can still be calculated by transforming it into a directed network by assuming a uniform distribution over edge weights.

To measure causal emergence, one needs to define a higher scale (some dimension reduction of the original system). In networks, macroscales are networks that are comprised of \textit{macro-nodes}. Each macro-node is a subgraph of micro-nodes that are grouped together and replaced with a single node, $\mu$. A macro-node $\mu$ has some $W^{out}_{\mu}$ that replaces the corresponding microscale outputs of nodes in the subgraph. We refer to a macroscale as any network with macro-nodes, wherein a network without macro-nodes is a microscale. Degrees are possible, however, as some networks may have many macro-nodes that increase $EI$, while other networks may have few.

Note that macroscales should generally be consistent, in that they will produce identical or approximately identical dynamics to those of the underlying microscale. In networks this can be assessed by comparing the dynamics of random walkers at each scale to see the extent to which the macroscale recapitulates the dynamics of the microscale. That is, to what degree the macroscale is \textit{consistent} with the underlying microscale in terms of its dynamics of random walkers. Different types of macro-nodes (constructed in different ways to summarize a subgraph's behavior) are sometimes needed to maintain consistency, such as using higher-order properties \cite{xu2016representing, klein2020uncertainty}. Here, all macro-nodes are of the $\mu | \pi$ type (based on the stationary distribution) since it has been shown that such nodes are in general highly consistent \cite{klein2020uncertainty}. The $W^{out}_{\mu|\pi}$ is a weighted average of each node in the subgraph's $W^{out}$, weighted by the stationary distribution, $\pi$, of the micro-nodes in the subgraph $S$ that constitute the new macro-node, $\mu$:

\begin{equation}\label{eq:ei}
   W_{\mu|\pi}^{out} = \displaystyle\sum_{i \in S} W_i^{out} \cdot \Big(\dfrac{\pi_i}{\sum_{k\in S} \pi_k}\Big)
\end{equation} wherein $\pi_i$ is the stationary probability of a random walker being on node $i$ in $S$ and $\Big(\dfrac{\pi_i}{\sum_{k\in S} \pi_k}\Big)$ is the summed stationary probability of a random walker being on one of the other nodes in the subgraph.

Finding macro-nodes that produce a gain in $EI$ can be thought of as an iterative procedure, wherein a subgraph is grouped in a macro-node and then the resulting change in $EI$ is calculated; this is followed by testing a new grouping of a subgraph, comparing its $EI$ to the original, and so on. For further details on the coarse-graining of networks, see \cite{klein2020uncertainty}. Below, we compare and detail algorithmic variants on this approach.

\subsection{\label{sec:algorithms}Algorithmic approaches to identifying causal emergence}

\subsubsection{\label{sec:greedyalgorithm}Greedy algorithm}

The greedy algorithm, which was first introduced in \cite{klein2020uncertainty}, is structured as follows: for each node, $v_i$, a list of neighboring nodes is constructed, $v_j \in B_i$, where $B_i$ is a subgraph of nodes ``surrounding'' node $v_i$; these include nodes with out-weights $w_{ji}$ connecting to $v_i$ (the ``parents'' of $v_i$), nodes that $v_i$'s out-weights $w_{ij}$ connect to (``children'' of $v_i$), and nodes that have out-weights $w_{kj}$ to $v_i$'s children (the ``parents of the children'' of $v_i$). Therefore $v_j \in B_i$ includes only those nodes targeting $v_i$, nodes targeted by $v_i$, and also the nodes that target those nodes targeted by $v_i$, a subgraph that is reminiscent to a \textit{Markov blanket}, from causal inference \cite{pearl2014probabilistic}. The algorithm assesses the change in $EI$ after a node $v_i$, and another node, $v_j \in B_i$, are combined into a macro-node, $v_\mu$. If this leads to a gain in $EI$, the algorithm stores this change. If necessary, it will change the queue of nodes, $v_j \in B_i$, adding any new neighboring nodes from $v_j$'s surrounding nodes, $B_j$, that were not already in $B_i$, so as to expand the search. If a node, $v_j$, has already been combined into a macro-node via a grouping with a previous node, $v_i$, then it will not be included in the new queue of nodes to check. Each pair of nodes is iteratively checked by the algorithm, starting with some node $v_j$, and pairing it with every node $v_j \in B_i$, and then starting on a new node, until every node is tested.

Given a network with $n$ nodes, checking a single pair of nodes for causal emergence requires computing a macroscale network, which has $O(n^2)$ time complexity, and then computing the $EI$ of the candidate macro network, $G_M$, which is also $O(n^2)$. In the worst case, $n\choose 2$ pairs of nodes need to be checked, though in practice interesting networks typically require far fewer checks, so the overall runtime of this algorithm is $O(n^4)$, though if optimized, this can be $O(n^3)$ by only recalculating and storing the change in $EI$ at every step. To use this greedy algorithm we made use of the publicly-available Python package at \url{github.com/jkbren/einet}.

\subsubsection{\label{sec:spectralanalysis}Spectral analysis}
Historically, spectral methods have been successful in obtaining partitions of graphs with desirable properties and good theoretical guarantees \cite{Fiedler, Guattery, SPIELMAN}. We use a novel variation of classical spectral algorithms to identify causal emergence accurately and efficiently.

Here we define some of the concepts relevant to this algorithm, for a full review of the related linear algebra, see \cite{strang09}. Given a graph $G$ with $n$ nodes, the \textit{adjacency matrix} $\mathbf{A}\in \mathbb{R}^{n \times n}$ of $G$ is the matrix where the entries $a_{ij}$ are equal to the weight of the edge from node $i$ to node $j$ in $G$ if this edge exists, and zero otherwise. The \textit{normalized adjacency matrix} is given by dividing the columns of $A$ by the degrees of the corresponding nodes, and can be thought of as the transition probability matrix of a random walk on $G$. The \textit{kernel} of a matrix $\mathbf{M} \in \mathbb{C}^{n \times n}$ is the set of vectors $\{\mathbf{v} \in  \mathbb{C}^{n}: \mathbf{Mv}=\mathbf{0} \}$, where $\mathbf{0}$ is the vector of all zeros. The kernel of $\mathbf{M}$ forms a linear subspace of $\mathbb{C}^{n }$.

Given the transition probability matrix $W_{out}$ of a network, our spectral algorithm calculates the eigendecomposition $\Lambda = \{\lambda_i\}$, $E=\{e_i\}$ of $W_{out}$, where $\lambda_i$ is the $i$th eigenvalue of $W_{out}$ and $e_i$ is the corresponding eigenvector. We obtain a basis $E'$ for the span of $W_{out}$ by removing the kernel and weighting the vectors by their associated eigenvalues: $E'=\{\lambda_i (e_i) \, | \, \lambda_i \neq 0 \}$. Intuitively, disregarding the kernel in our analysis makes sense because it corresponds to degeneracy in the corresponding network. Therefore, considering the span gives us a description of the network topology without the components that generate degeneracy. Additionally, the nonzero eigenvalues and corresponding eigenvectors of $W_{out}$ contain rich information about the topological structure of the network. 

Notably, we find that using the full non-degenerate spectrum of eigenvalues and corresponding eigenvectors rather than using a few of the most significant, as is common in other methods such as those discussed below, significantly improves the algorithm's ability to discover good macro-nodes.

We use $E'$ to associate each node $v_j$ in a network with a vector composed of the entry in each eigenvector corresponding to $v_j$. We calculate a distance metric for all pairs of nodes in the network by taking the cosine similarity of these vectors. If a pair of nodes are not in each other's surrounding subgraphs, $B_j$ and $B_i$, then grouping them together cannot increase $EI$, so we define the distance between them to be $\infty$. We then apply the OPTICS clustering algorithm \cite{OPTICS} to this distance matrix to obtain a clustering over the nodes of the network, the output of which we interpret as a mapping from the microscale to the new macroscale. The quality of this coarse-graining (i.e., the amount of causal emergence it discovers) depends on the distance threshold $\epsilon$ used in the clustering, and the optimal value for $\epsilon$ depends on the topology of the network and is difficult to select \textit{a priori}. Therefore we check the \textit{EI} gain over a range of $\epsilon$ values to find the best clustering, which can be done efficiently by the OPTICS algorithm.

Our proposed algorithm is related to the spectral coarse-graining algorithm proposed by \cite{Gfeller2007, gfeller2008spectral}. Here we use a similar procedure, but explicitly optimize the \textit{EI} of the coarse-grained network, while the aforementioned algorithm seeks only to find a coarse-graining that preserves the spectral characteristics of the original network. We do not directly seek to preserve the spectral characteristics, but by preserving the behavior of random walkers this happens implicitly, since the dynamics of random walkers are determined by spectral properties. In particular, our approach differs from this previous algorithm by using the network's full set of non-degenerate eigenvectors, employing a more sophisticated clustering procedure, and considering a range of possible coarse-grainings by checking a range of different $\epsilon$ values in the clustering step, all of which were found to improve the algorithm's ability to find \textit{EI}-maximizing coarse grainings.

Regarding algorithmic complexity, given a network with $n$ nodes, eigendecomposition can be performed in $O(n^3)$, computing the OPTICS reachability graph is $O(n \log n)$, and computing the clustering for a given $\epsilon$ is $O(n)$. Only a constant number of different $\epsilon$'s are considered, so the overall complexity is $O(n^3)$, though in practice the runtime is dominated by computing the OPTICS reachability graph for networks at least up to $5000$ nodes. Additionally, both the eigendecomposition and clustering operations are parrallelizable.

We speculate that there is a deep connection between the kernel of the adjacency matrix and the scale of the corresponding network, and this is why analyzing the adjacency matrix works for finding causal emergence. For example, Erdős–Rényi random graphs exhibit essentially no causal emergence, and almost always have kernel dimension zero \cite{Chung}. Likewise, star graphs are optimal at the macroscale and exhibit maximal causal emergence, and a star graph of size $n$ has kernel dimension $n-2$ \cite{Seidy}. Comparatively, both star graphs and typical Erdős–Rényi graphs have Laplacian kernel dimension zero.

The presence of macroscales in a given network is determined by the amount of indeterminism and degeneracy present in the network. Using tools from linear algebra, here we derive a connection between the basic properties of adjacency matrices and the $EI$ of networks that motivates the use of spectral analysis for determining scale.

Degeneracy is an indication of attractor dynamics in a system---it measures the number of states which converge onto the same state in the future, a phenomena which we claim can also be quantified using the algebraic properties of adjacency matrices. Given a network $G$ with $n$ nodes, let $\mathbf{x_t}$ be a vector representing a distribution of random walkers on a graph $G$: $\mathbf{x_t}[i]=\Pr(\textit{walk is on node i at time t})$; note that $\mathbf{x_t}$ could represent a deterministic state if all components except for one are zero. Let $\mathbf{A_{norm}}$ be the degree-normalized adjacency matrix of $G$, then the distribution at the next time step is given by $\mathbf{x_{t+1}}=\mathbf{A_{norm}}\mathbf{x_t}$. In order to quantify degenerate behavior, we can use the kernel of $\mathbf{A_{norm}}$ to construct distributions $\mathbf{w}$ that differ from $\mathbf{x_t}$ and also transition to $\mathbf{x_{t+1}}$, but care is required to ensure that $\mathbf{w}$ is a valid probability distribution. As such, let $\mathbf{v}$ be any real vector in the kernel of $\mathbf{A_{norm}}$ where $\mathbf{v}[i]$ is non-negative if $\mathbf{x_t}[i]$ is zero, and let $\beta=\max\limits_{i \leq n} \frac{\mathbf{x_t}[i]}{\mathbf{v}[i]-\mathbf{x_t}[i]}$. Then for any $b \in [0,\beta]$ we can construct $\mathbf{w}=\mathbf{x_t} + b\mathbf{v}$ satisfying $\mathbf{A_{norm}}\mathbf{w} =  \mathbf{A_{norm}}(\mathbf{x_t}+b\mathbf{v})=\mathbf{A_{norm}}\mathbf{x_t}+b\mathbf{A_{norm}}\mathbf{A_{norm}}=\mathbf{A_{norm}}\mathbf{x_{t+1}} + \mathbf{0}=\mathbf{A_{norm}}\mathbf{x_{t+1}}$. Scaling $\mathbf{v}$ by $b$ ensures that the entries of $\mathbf{w}$ are non-negative. By definition, $\mathbf{x_t}$ is a valid probability distribution, so its components sum to 1. Multiplication of vectors by Markov matrices preserves the sum of their components, and since $\mathbf{Av}=\mathbf{0}$, the components of $\mathbf{v}$ must also add up to 0. Thus we have $\sum_{i=1}^n \mathbf{w}[i] = \sum_{i=1}^n \mathbf{x_t}[i]+ b\sum_{i=1}^n \mathbf{v}[i] = 1 + 0$, and we can conclude that $\mathbf{w}$ is a valid probability distribution. If we additionally assume that all components of $\mathbf{x_t}$ are nonzero, the set of such degenerate distributions $\mathbf{w}$ forms a convex region of a linear subspace of the same dimension as the kernel of $\mathbf{A_{norm}}$. Conversely, if we have two distributions $\mathbf{x_t^1}$ and $\mathbf{x_t^2}$ that both transition to $\mathbf{x_{t+1}}$, then the vector $\mathbf{x_t^1}-\mathbf{x_t^2}$ is in the kernel of $\mathbf{A_{norm}}$: $\mathbf{A_{norm}}(\mathbf{x_t^1}-\mathbf{x_t^2})=\mathbf{A_{norm}}\mathbf{x_t^1} - \mathbf{A_{norm}}\mathbf{x_t^2}=\mathbf{x_{t+1}}-\mathbf{x_{t+1}}=\mathbf{0}$.

This tells us that degenerate dynamics in a network correspond directly with the kernel dimension of the adjacency matrix. For any arrangement of degenerate nodes in $G$, there is a corresponding element in the kernel of $\mathbf{A_{norm}}$, and for every element in the kernel of $\mathbf{A_{norm}}$, there is a set of degenerate nodes. Therefore, we conclude that there is a strong relationship between the connectivity and algebraic structures of graphs, and the effectiveness of spectral analysis for determining scale is a result of this connection.

\subsubsection{\label{sec:gradient}Gradient descent}

Gradient descent is a powerful approach for solving a wide range of optimization problems, and it is a ubiquitous approach in machine learning \cite{ruder2016overview, nesterov1983method}. However, it is not immediately applicable to the problem of finding good macroscale networks, because we need to optimize $EI$ as a function of a set partition (coarse-graining), but this function is not differentiable, which is a requirement for performing gradient descent. Given a network with $n$ nodes, we relax the problem by replacing the set partition with a matrix $M \in \mathbb{R}^{n \times n}$ with entries $m_{i\mu}=\Pr(v_i \in v_\mu)$ for micro-node, $v_i$, and macro-node, $v_\mu$. Intuitively, the entries $m_{i\mu}$ represent how confident the algorithm is that node $i$ should be placed in macro-node $\mu$ in order to maximize the $EI$ at the macro-scale. For the purposes of optimization, this matrix is represented as unconstrained real numbers, to which the softmax function $\sigma(\mathbf{x}) = \exp{x_i}/\sum_{j=1}^k \exp{x_j}$ for $\mathbf{x}=(x_1,...,x_k) \in \mathbb{R}^k $ and $i = 1,...,k$ is applied to normalize the the columns of \textit{M} into valid probability distributions. Using this relaxation, we can compute the $EI$ of the ``probabilistic'' coarse graining as a differentiable function of $M$ and the network adjacency matrix, allowing $M$ to be optimized using gradient descent with momentum. Specifically, we obtain the candidate macroscale network by multiplying the microscale adjacency matrix by $M$ and then we compute the $EI$ of the macroscale network as a function of $M$ and the network adjacency matrix, compute the gradient of the $EI$ with respect to $M$, and use this gradient to maximize the $EI$ using a standard gradient descent algorithm. $M$ is initialized randomly, and updated until convergence, or up until a certain number of iterations. When it converges, the result is a ``deterministic'' coarse graining which can be interpreted in the same way as the outputs of the previous two algorithms. A disadvantage of this approach is that the convergence of $M$ depends on the random initialization, so multiple runs on the same network may produce different results. Performance also depends on the learning rate used in gradient descent, and the maximum number of iterations allowed.

The time complexity of a single iteration of the gradient descent algorithm is dominated by a constant number of matrix multiplications, where the matrix sizes correspond to the size of the network being coarse-grained, so a single iteration can be done naively in $O(n^3)$. Since at most only a constant number of iterations are performed (while in practice this constant is large), the overall time complexity of this approach is also $O(n^3)$. While the performance of this algorithm is asymptotically equivalent to that of the spectral approach, in practice the spectral algorithm is much faster on all instances of reasonable size.

\begin{figure}[t!]
    \centering
    \includegraphics[width=\textwidth]{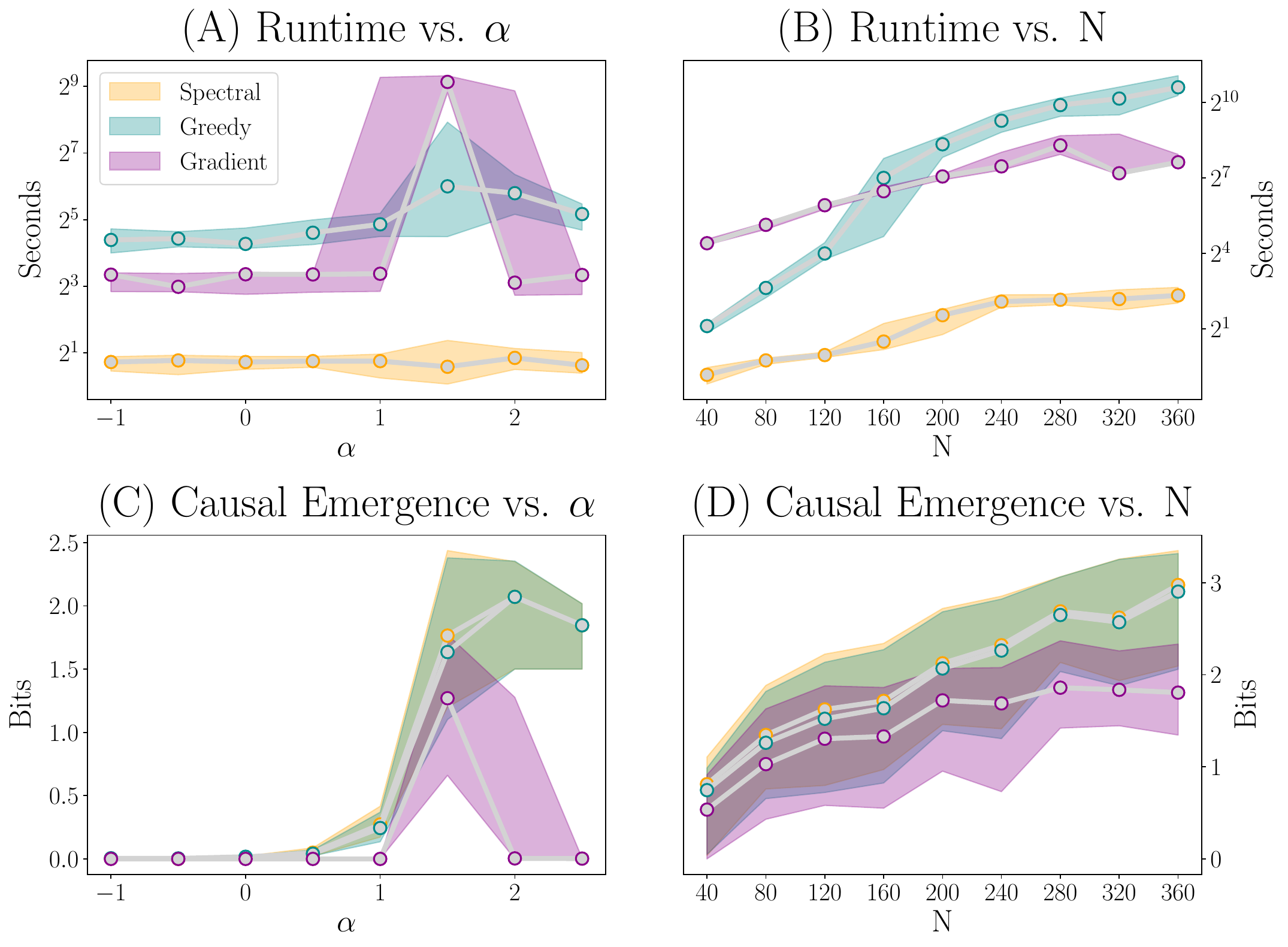}
    \caption{\textbf{Comparison of Methods for Computing Causal Emergence \textbf(A)} Comparing algorithm runtimes (in seconds) for different values of $\alpha$, calculated by adding the system and user CPU time used by the process during calls to each function. Both the greedy and gradient descent algorithms struggle with detecting mesoscales around $\alpha=1.5$ while the spectral algorithm remains relatively constant. \textbf{(B)} Runtime vs. network size. While the greedy algorithm is more efficient than the gradient descent algorithm on small instances, gradient descent scales better to larger networks. The spectral algorithm outperforms both. \textbf{(C)} Causal emergence (in bits) found by different algorithms for preferential attachment networks with different values of $\alpha$. The greedy and spectral algorithms behave similarly, while gradient descent struggles in the more subtle $\alpha<1$ cases. \textbf{(D)} Detected causal emergence (in bits) compared to network size, $n$. The spectral algorithm slightly outperforms the greedy algorithm, and gradient descent lags behind both.}
    \label{fig2:algos}
\end{figure}

\section{\label{sec:results}Results}

\subsection{\label{sec:comparingmethods}Comparing methods of finding macroscales}

What is the best way to find the scale at which $EI$ is high, along with the associated changes to network properties? A brute-force search is impossible due to the number of subgraphs (the same as the number of partitions). However, the challenge faced is similar to the challenge of finding communities of nodes, which is common in network science, even though it too scales by the number of partitions. Here we compare three methods for finding $EI^{max}$: a greedy algorithm (described in Section \ref{sec:greedyalgorithm}), a method based on gradient descent (Section  \ref{sec:gradient}), and a method based on spectral analysis (Section  \ref{sec:spectralanalysis}).

First, we analyzed the computational runtime of these different algorithms for networks of 150 nodes at varying degrees of preferential attachment (Fig. 1A). Preferential attachment networks were used because they provide a range of connectivities controlled by $\alpha$, which indicates the degree of preferential attachment. When $\alpha$ = 1 attachment leads to a scale-free network, whereas $\alpha$ approaches 0 this generates trees with long path lengths, and when $\alpha > 2$ this leads to hub-and-spoke models. Each node adds $m$ edges connect to nodes already in the network, $v_j$, with a probability proportional to $k_j^\alpha$ \cite{Krapivsky2001}. Unless otherwise specified $m = 1$ in our simulations.
What we noticed is that extreme microscales (minimal to no causal emergence) and extreme macroscales (high levels of causal emergence) do not require much computational resources to discover. These can be viewed as cases where reduction or emergence are very clear based on the system architecture. However, networks with significant mesoscales ($1.0 < \alpha < 2.0$) require significant runtime. Notably, this is less so for the spectral analysis, and also the runtime is several orders of magnitude lower in all conditions. This is true even when the node number of the network is significantly increased (Fig. 1B).

Next we examined whether the algorithms successfully captured causal emergence and found informative higher scales (Fig. 1C). While all algorithms could identify cases of causal emergence, both the spectral and the greedy seemed to perform better and find equivalent cases of causal emergence. This pattern continued even when the number of nodes was increased (Fig. 1D).

\subsection{\label{sec:macroscales}Network properties of macroscales in preferential attachment networks.}

\begin{figure}[t!]
    \centering
    \includegraphics[width=\textwidth]{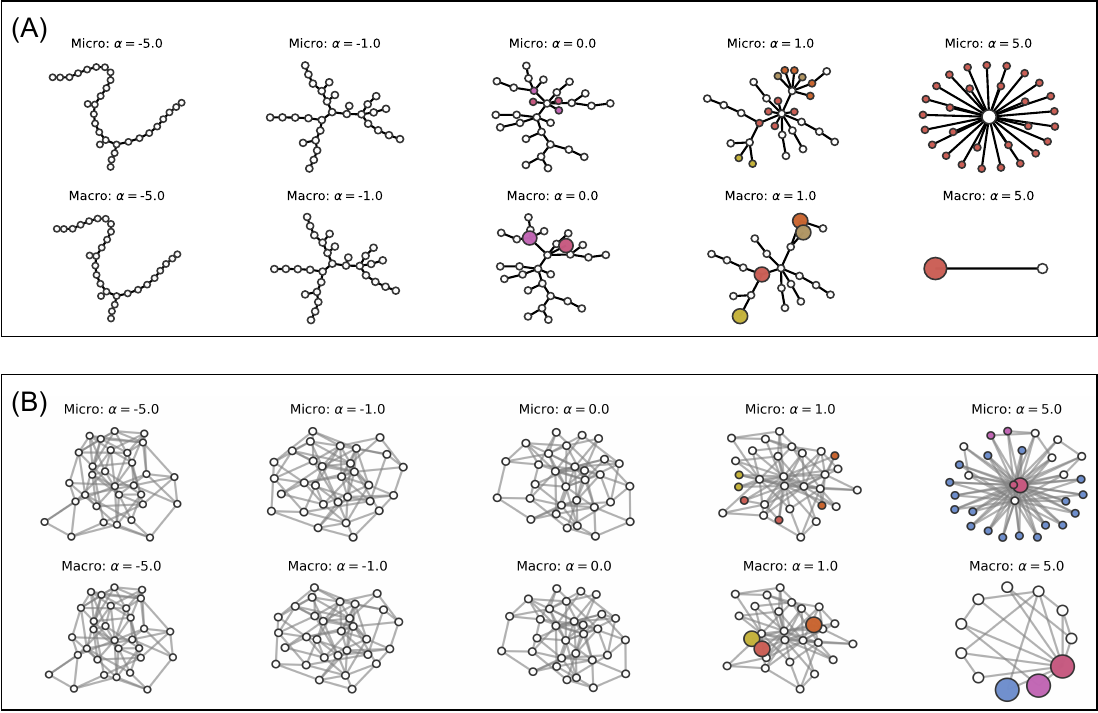}
    \caption{\textbf{Example macroscales of networks \textbf(A)} Preferential attachment networks of 50 nodes (\textit{m} = 1) are shown with increasing $\alpha$ (top). Our spectral algorithm is used to identify macro-nodes, which are also shown (bottom). Nodes that are grouped into macro-nodes by the algorithm are color coded. \textbf{(B)} Preferential attachment networks but with \textit{m} = 3, meaning the networks are denser as more edges are added each step. Informative macro-nodes also occur at higher levels of $\alpha$.}
    \label{fig2:algos}
\end{figure}

To explore how to find informative higher scales of networks, we repeatedly simulate networks grown under preferential attachment rules \cite{barabasi2009scale}. In a preferential attachment growth model, the network grows each time step by adding a new node with its $m$ new edges to the network (a process described in previous section). Our reasoning for using preferential attachment network is twofold: first, just by manipulating $\alpha$ one can span a range of different network connectivities in a controlled manner. Second, preferential attachment networks can span the range of causal emergence, from causal reduction (when no $EI$ is gained by dimension reductions, occuring when $\alpha < 1$), to when all nodes are grouped into a single macro-nodes (which occurs $\alpha > 3$).

Using these networks, we demonstrate the results of our proposed spectral algorithm in Section \ref{sec:effectiveinformation}. In Fig. 2A both the original microscale of 50-node networks (top) are shown, with nodes colored depending on what macro-node they the spectral algorithm groups them into, as well as the resultant macro-nodes identified by the algorithm (with each macro-node color corresponding to the colors of nodes grouped into them). As can be seen, within a significant domain of preferential attachment causal emergence cannot happen to any real degree, for example, if $\alpha < 1.0$. This corresponds to the region of sublinear preferential attachment, in which a network does not develop higher scales. It is notable that once preferential attachment is superlinear, and therefore no longer ``scale-free'' according to traditional network science, causal emergence becomes significant. As $\alpha$ increases, the number of nodes grouped into the macro-nodes increases until it is $n-1$ at high levels of $\alpha$ (see \cite{klein2020uncertainty} for more details.) Our spectral analysis is successful in finding macro-nodes, and identifying a similar pattern of increasing causal emergence with higher levels of $\alpha$, even in the less differentiated and non-tree networks where \textit{m} = 3 (Fig. 2B).

Next we explore how macroscales that maximize $EI$ (Fig. 3A) also change in their other network properties beyond just $EI$, here making use of larger preferential attachment networks (\textit{n} = 150, \textit{m} = 1). As defined in \cite{klein2020uncertainty}, $EI = \text{determinism} - \text{degeneracy}$. Both determinism and degeneracy define how random walkers move about the network. Specifically, $determinism$ is based on how much information is not lost to indeterminism, that is, a walker's uncertainty while on the average node. The indeterminism is $\langle {H}(W^{out}_{i}) \rangle$, and the determinism, its inverse, is $\log2(n) - indeterminism$. This is because $\log2(n)$ represents the upper bound of $indeterminism$ which only occurs when all walkers face only $w_{ij} = 1$, i.e., when all walkers move deterministically no matter what node they are on. Meanwhile, the degeneracy of a network describes how the weight of a network is distributed. If all nodes lead only to one node, that network is perfectly degenerate. The degeneracy can be captured by $log_2(n) - H(\langle W^{out}_{i} \rangle)$ which is a measure of how non-uniform the weight distribution is of the network. The increase in determinism and decrease in degeneracy is the cause of the increase in $EI$ at the macroscale. As can be seen in Fig. 3B, the indeterminism of causally-emergent macroscale networks decreases, and Fig. 3C shows the decrease in degeneracy at the macroscale. This means that random walkers face less uncertainty at the macroscale of the network. 

In Fig. 3D, we show how the entropy rate of random walkers is much lower at a causally-emergent macroscale compared to its microscale. As defined in \cite{xu2016representing}, the entropy of random walkers on the network is $H{\big(}X_{t+1}|X_{t} = \sum_{i,j}\pi iw_{ij}\log_2 w_{ij}\big)$, where $\pi(i)$ is the stationary probability, and $w_{ij}$ is the probability of $v_j$ given $v_i$ for a random walker (assuming normalization such that edges are equivalent to probabilities of random walks). Note that, the "reverse" entropy rate (the entropy derived if the random walkers are walking backwards across the network) shows a similar reduction at the macroscale (not shown due to its similarity with the entropy rate).

Global network efficiency is a measure of network connectedness based on path length, often used to quantify small-world network dynamics \cite{PhysRevLett.87.198701}. The global efficiency of a network, $G$, is the inverse of the average shortest path between all pairs of nodes in a network. The global network efficiency can also be greater at macroscales in cases of causal emergence (Fig. 3E). This indicates that macro-node communication or interaction is more efficient than their underlying micro-scale, as the average path length is much lower. The betweenness centrality of a node identifies how many shortest paths of the network traverse that node \cite{freeman,barabasi2016network}. Nodes that receive many edges are likely to have a high centrality and thus exert control over network dynamics. A higher betweenness centrality is commonly used to measure the importance of nodes for information transmission in a network, in cases ranging from telecommunications  \cite{onnela2007structure, zhang2014finding} to gene regulatory networks  \cite{koschutzki2008centrality} to even networks of scientific citations \cite{leydesdorff2007betweenness}. Comparing the centrality of $G$ to $G_m$ we see that that the average centrality of the network's nodes increases significantly after causal emergence (Fig. 3F). Since betweenness centrality can be interpreted as quantifying how much control nodes in the network have over information transmission, this indicates that macroscales have a higher degree of control over dynamics than their underlying microscales. Another common notion of centrality is the \textit{eigenvector centrality} of nodes in a network, which not only corresponds to the degree of a given node but also considers nodes whose neighboring nodes have a high degree \cite{ruhnau2000}. In Fig. 3G, we again see that the average eigenvector centrality scores for the macro-nodes begins to increase as $\alpha$ increases.

The communicability of a network is a generalization of the shortest path between any two nodes in a network \cite{estrada}, and the entropy of the communicability sequence has recently been used to characterize and compare networks \cite{chen2018}. We show in Fig. 3H the behavior of the communicability sequence entropy as $\alpha$ increases, showing that the communicability sequence of the macro-nodes begins to decrease after $\alpha=1.0$, suggesting that the macro-nodes' communicability with other nodes in the network becomes less uniform (i.e., they have higher determinism).

Creating macro-nodes has a slight, though increasingly negative, effect on the node degree as $\alpha$ increases (Fig. 3I), but we really see the effect of macro-nodes when we observe the variance of the degree distribution (Fig. 3J). The detection of macro-nodes in a preferential attachment network dramatically increases the variance of the degree of the remaining micro-nodes. Similarly, we see the network becoming more degree disassortative as $\alpha$ increases, which is more pronounced in the micro-nodes than in the macro-nodes. 

Finally, as discussed in Section \ref{sec:spectralanalysis}, the kernel dimension of the network changes as $\alpha$ increases, with the kernel dimension of the micro-nodes increasing rapidly, while the macro-nodes decrease (Fig. 3L). This behavior offers insights into the possible mechanisms behind why the spectral approach to discovering causal emergence in networks is as effective as it appears to be. There is an inherent separation between nodes in the microscale and macroscale, which the spectral algorithm uses to inform which nodes to partition into macro-nodes.

These extensive and significant changes to network properties show how and why the efficient identification of macroscales in networks is important, such as through our proposed spectral algorithm. Notably, these properties change despite the macroscale and microscale still be consistent in terms of their random walk dynamics, indicating that network properties are in general dependent on scale.

\begin{figure}[t!]
    \centering
    \includegraphics[width=\textwidth]{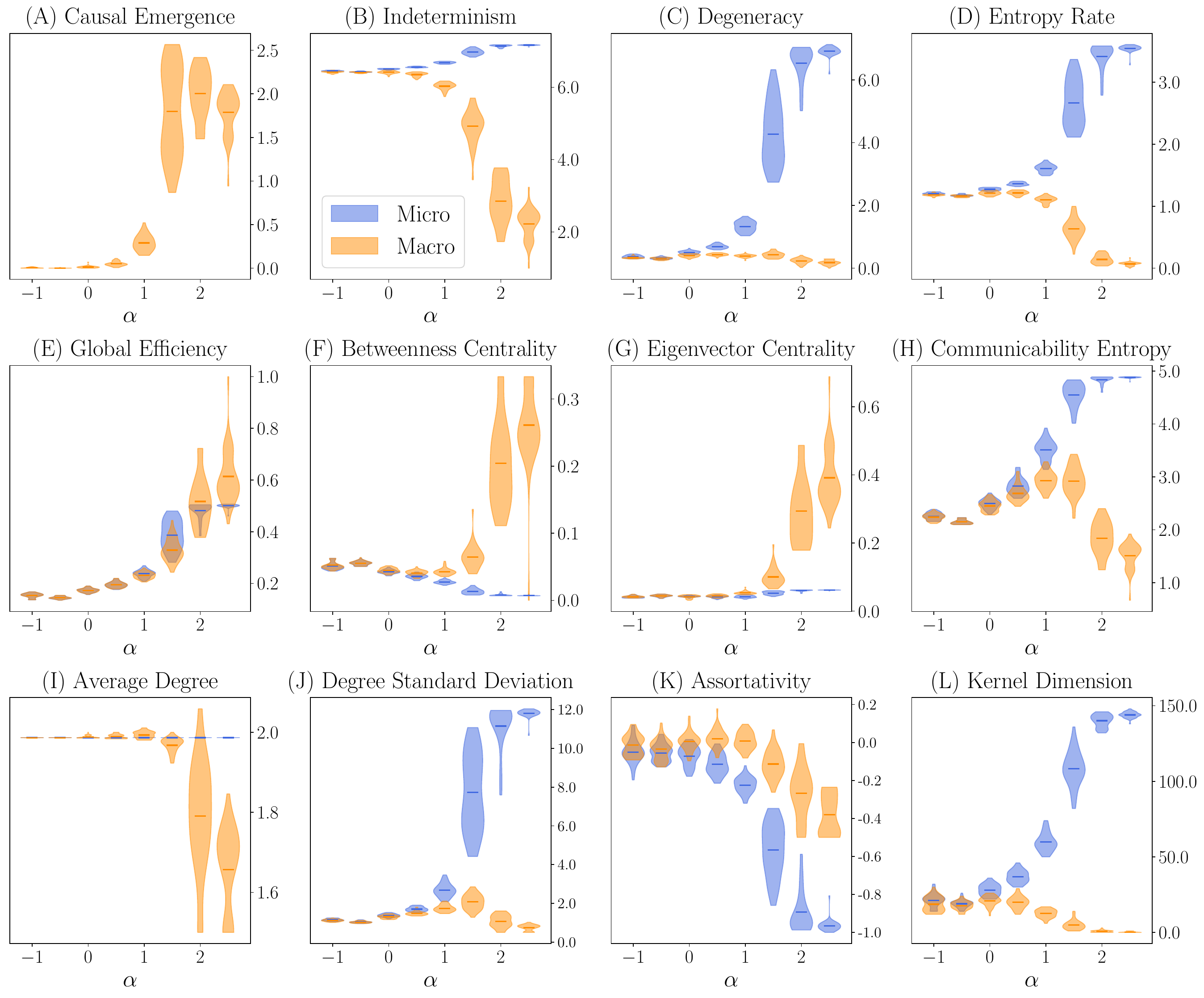}
    \caption{\textbf{Preferential attachment network properties at different scales.} All macroscales shown here were obtained using the spectral algorithm.
    \textbf{(A)} The amount of causal emergence found in 150 node preferential attachment networks for different values of $\alpha$, calculated using a sample of 40 graphs per value of $\alpha$. For small $\alpha$ the microscale gives a near optimal description. After $\alpha=1.0$ mesoscales emerge, and above $\alpha=2.0$, the macroscale dominates.
    \textbf{(B)} Determinism of node outputs increases at higher scales. 
    \textbf{(C)} Similarly, networks are less degenerate (sharing targets) at the macroscale.
    \textbf{(D)} The entropy rate of random walkers over the stationary distribution of the network at both micro- and macro-scales.
    \textbf{(E)} Networks with a distinct mesoscale around $\alpha=1$ are slightly more efficient at the microscale, but once the macroscale becomes dominant for $\alpha > 1.5$ the macro network is much more efficient.
    \textbf{(F)} Betweenness centrality is a measure of how important individual nodes are to the overall topology of the network.
    \textbf{(G)} Eigenvector centrality of the resulting macronodes increases as $\alpha$ gets larger.
    \textbf{(H)} Entropy of the communicability sequence of nodes in the network, with differences between the microscale and macroscale emerging at $\alpha \approx 1.0$
    \textbf{(I)} Average degree of the network decreases as more macronodes emerge (via the rapid reduction in degenerate links).
    \textbf{(J)} The variance of the node degree increases as $\alpha$ increases. 
    \textbf{(K)} In general, the degree assortativity decreases as $\alpha$ increases, which corresponds to the emergence of more disassortative. However, the macroscale assortativity decreases less than the nodes in the microscale.
    \textbf{(L)} The kernel dimension relates to the degeneracy of a network. From a probabilistic perspective, it quantifies the space of distributions of random walkers that converge to the same distribution from one timestep to the next.}
    \label{fig3:metrics}
\end{figure}

\section{\label{sec:discussion}Discussion}

Networks can possess macroscales that often have different network properties than their underlying microscales. Finding informative higher scales is a search procedure very similar to community detection, as it is sorting through the number of possible partitions of the network. While in this sense finding macro-nodes is similar to identifying communities of nodes, generally community detection is focused on subgraphs that have more in-group connectivity than out-group \cite{leskovec2010empirical}. Comparatively, macro-nodes represent subgraphs that possess a viable summary statistic in terms of their behavior in the network, specifically, they seek to preserve the behavior of random walkers while at the same time decreasing the uncertainty of their behavior. Since what matters in their formation is that random walk behavior is preserved, macro-nodes can be over a range of connectivity. Additionally, after finding appropriate partitioning into subgraphs, macro-nodes are a recasting of the network itself via a transformation of those subgraphs into individual nodes.

In order to find informative higher scales we used similar metrics and methods as we previously introduced to identify higher scales of networks via causal emergence \cite{klein2020uncertainty}. Specifically, causal emergence is the degree to which effective information (the amount of information in the connectivity of a network), increases at higher scales. Identifying causal emergence in networks has numerous benefits in terms of reducing the dimension of networks, but also improving various network properties, from the structural to the informational. Macroscales that are causally emergent can show, in relation to their original microscale: a decrease in indeterminism, a decrease in degeneracy, a decrease in the entropy rate of random walkers at the stationary distribution, an increase in global network efficiency, an increase in average betweenness centrality, and a decrease in the kernel dimension. Therefore, it is important that network scientists, if they wish to understand the function and structure of a network, explicitly model higher scales. This requires finding them.

We therefore compared three algorithms: one based on gradient descent, another on a greedy approach, and the other on spectral analysis. An evolutionary algorithm was also attempted but its performance was significantly worse (both in finding causal emergence and in computational time) that we did not include it in the results. Upon comparing the search methods, we can pinpoint the space of connectivity that is most difficult, but also rewarding, to find higher scales at. This is over the mesoscale of the network wherein the higher scale is composed of a complex array of macro-nodes of different sizes. Notably, this indicates that the systems that are most difficult to find an informative scale for possess mesoscales, which may explain the difficulties in understanding the functional architecture of complex systems like the brain
\cite{buxhoeveden2002minicolumn, yuste2015neuron}.

Of the three investigated algorithms we found that spectral analysis performed orders of magnitude better in terms of runtime and got equal to or better results in terms of identifying cases of causal emergence in preferential attachment networks. Therefore, we recommend those that want to find informative higher scales use a clustering algorithm; specifically, a modified form of spectral clustering introduced in \ref{sec:spectralanalysis} shows promise for larger networks of thousands of nodes. For this purpose, this spectral analysis function was added to the publicly-available Python package for calculating causal emergence at \url{github.com/jkbren/einet}. The algorithms were here only tested in preferential attachment networks under different growth rules. While this means our results are restricted to these types of networks, it is worth noting that preferential growth creates sparse networks of the real-world sort seen in nature and technology \cite{barabasi1999emergence}. It should be noted that there can be divergences between real-world networks and that modifications to pure preferential attachment might be needed to account for real-world network structure \cite{kaiser2004spatial, shang2017fitness}, which could limit the application of our results. However, there is already evidence that the spectral analysis techniques works well in finding causal emergence in real-world networks based on protein interactomes from over 1000 species \cite{hoel2020evolution}. In future research we would like to explore architectures different than the range of preferential attachment networks we have here considered, further ground the relationship between spectral analysis and grouping, as well as apply these techniques to more real-world networks.

Overall, our results indicate that network macroscales can be more informative in terms of a higher $EI$, but also in terms of measures like the entropy rate. As the topology of a network changes across scales, its network properties change as well, often in ways that suggest that networks that possess informative higher scales should be understood as operating at that scale, since macroscales can entail a peak of efficiency, centrality, or information transmission above and beyond the underlying microscale.

\section{\label{sec:data}Data availability statement}
The code used for the findings of this study are openly available at \url{github.com/jkbren/einet}.

\section{\label{sec:acknowledgments}Acknowledgments}
\textbf{Funding:} This work was supported by Army Research Office Grant W911NF2010243. It was also supported by a grant from Templeton World Charity Foundation, Inc. (TWCFG0273). \textbf{Author contributions:} R.G., B.K., and E.H. conceived the project. R.G., B.K., and E.H. wrote the article. R.G. performed the analyses. \textbf{Competing interests:} The authors declare no competing interests.

\printbibliography[title={References}]

@book{barabasi2016network,
  title={Network Science},
  author={Barab{\'a}si, Albert-L{\'a}szl{\'o}},
  year={2016},
  publisher={Cambridge University Press},
  isbn={9781107076266},
}

@article{buxhoeveden2002minicolumn,
  title={The minicolumn hypothesis in neuroscience},
  author={Buxhoeveden, Daniel P and Casanova, Manuel F},
  journal={Brain},
  volume={125},
  number={5},
  pages={935--951},
  year={2002},
  doi = {10.1093/brain/awf110},
  publisher={Oxford University Press}
}

@article{yuste2015neuron,
  title={From the neuron doctrine to neural networks},
  author={Yuste, Rafael},
  journal={Nature reviews neuroscience},
  volume={16},
  number={8},
  pages={487},
  year={2015},
  doi={10.1038/nrn3962},
  publisher={Nature Publishing Group}
}

@inproceedings{leskovec2010empirical,
  title={Empirical comparison of algorithms for network community detection},
  author={Leskovec, Jure and Lang, Kevin J and Mahoney, Michael},
  booktitle={Proceedings of the 19th international conference on World wide web},
  ibsn={978-1-60558-799-8},
  pages={631--640},
  year={2010},
  organization={ACM}
}

@article{barabasi2009scale,
  title={Scale-free networks: a decade and beyond},
  author={Barab{\'a}si, Albert-L{\'a}szl{\'o}},
  journal={science},
  volume={325},
  number={5939},
  pages={412--413},
  doi={10.1126/science.1173299},
  year={2009},
  publisher={American Association for the Advancement of Science}
}

@article{xu2016representing,
  title={Representing higher-order dependencies in networks},
  author={Xu, Jian and Wickramarathne, Thanuka L and Chawla, Nitesh V},
  journal={Science Advances},
  volume={2},
  number={5},
  pages={e1600028},
  doi = {10.1126/sciadv.1600028},
  year={2016},
  publisher={American Association for the Advancement of Science}
}

@article{hoel2017map,
  title={When the map is better than the territory},
  author={Hoel, Erik},
  journal={Entropy},
  volume={19},
  number={5},
  pages={188},
  doi = {10.3390/e19050188},
  year={2017},
  publisher={Multidisciplinary Digital Publishing Institute}
}

@article{gallagher1999beyond,
  title={Beyond reductionism},
  author={Gallagher, Richard and Appenzeller, Tim},
  journal={Science},
  volume={284},
  number={5411},
  pages={79--80},
  year={1999},
  doi={10.1126/science.284.5411.79},
  publisher={American Association for the Advancement of Science}
}

@book{pearl2014probabilistic,
  title={Probabilistic Reasoning in Intelligent Systems: Networks of Plausible Inference},
  author={Pearl, Judea},
  year={2014},
  isbn={9781558604797},
  publisher={Elsevier}
}

@article{hoel2013quantifying,
  title={Quantifying causal emergence shows that macro can beat micro},
  author={Hoel, Erik and Albantakis, Larissa and Tononi, Giulio},
  journal={Proceedings of the National Academy of Sciences},
  volume={110},
  number={49},
  pages={19790--19795},
  year={2013},
  doi = {10.1073/pnas.1314922110},
  publisher={National Acad Sciences}
}

@article{hoel2016can,
  title={Can the macro beat the micro? Integrated information across spatiotemporal scales},
  author={Hoel, Erik and Albantakis, Larissa and Marshall, William and Tononi, Giulio},
  journal={Neuroscience of Consciousness},
  volume={2016},
  number={1},
  year={2016},
  doi={10.1093/nc/niw012},
  publisher={Oxford University Press}
}

@article{klein2020uncertainty,
  title={The emergence of informative higher scales in complex networks},
  author={Klein, Brennan and Hoel, Erik},
  journal={Complexity},
  url={https://arxiv.org/abs/1907.03902},
  year={in press}
}

@article{Fiedler,
author = {Miroslav Fiedler},
journal = {Banach Center Publications},
language = {eng},
number = {1},
pages = {57-70},
title = {Laplacian of graphs and algebraic connectivity},
url = {http://eudml.org/doc/267812},
volume = {25},
year = {1989},
}

@inproceedings{Guattery,
  title={On the performance of spectral graph partitioning methods},
  author={Stephen Guattery and Gary L. Miller},
  booktitle={SODA},
  year={1995}
}

@article{SPIELMAN,
title = "Spectral partitioning works: Planar graphs and finite element meshes",
journal = "Linear Algebra and its Applications",
volume = "421",
number = "2",
pages = "284 - 305",
year = "2007",
note = "Special Issue in honor of Miroslav Fiedler",
issn = "0024-3795",
url = {10.1016/j.laa.2006.07.020},
author = "Daniel A. Spielman and Shang-Hua Teng"
}

@INPROCEEDINGS{OPTICS,
    author = {Mihael Ankerst and Markus M. Breunig and Hans-peter Kriegel and Jörg Sander},
    title = {OPTICS: Ordering points to identify the clustering structure},
    booktitle = {Proc. ACM SIGMOD’99 Int. Conf. on Management of Data},
    year = {1999},
    doi={10.1145/304182.304187},
    pages = {49--60},
    publisher = {ACM Press}
}

@article {Chung,
	author = {Chung, Fan and Lu, Linyuan and Vu, Van},
	title = {Spectra of random graphs with given expected degrees},
	volume = {100},
	number = {11},
	pages = {6313--6318},
	year = {2003},
	publisher = {National Academy of Sciences},
	issn = {0027-8424},
	doi = {10.1073/pnas.0937490100},
	journal = {Proceedings of the National Academy of Sciences}
}

@article{Seidy,
  title={Spectra of some simple graphs},
  author={El Seidy,Essam and Eldin Hussein, Salah and AboElkher, Atef},
  year={2015},
  journal = {Mathematical Theory and Modeling},
  volume = {5},
  issue = {2},
  doi = {10.14419/ijamr.v5i2.6106},
  pages = {115-121}
}

@article{PhysRevLett.87.198701,
  title = {Efficient behavior of small-world networks},
  author = {Latora, Vito and Marchiori, Massimo},
  journal = {Phys. Rev. Lett.},
  volume = {87},
  issue = {19},
  pages = {198701},
  numpages = {4},
  doi={10.1103/PhysRevLett.87.198701},
  year = {2001},
  month = {10},
  publisher = {American Physical Society},
}

@inproceedings{nesterov1983method,
  title={A method for unconstrained convex minimization problem with the rate of convergence O (1/k\^{} 2)},
  author={Nesterov, Yurii},
  booktitle={Doklady AN USSR},
  volume={269},
  pages={543--547},
  year={1983}
}

@article{ruder2016overview,
  title={An overview of gradient descent optimization algorithms},
  author={Ruder, Sebastian},
  journal={arXiv:1609.04747},
  url={https://arxiv.org/abs/1609.04747},
  year={2016}
}

@book{strang09,
  author = {Strang, Gilbert},
  isbn = {9780980232714},
  publisher = {Wellesley-Cambridge Press},
  refid = {298178914},
  title = {Introduction to Linear Algebra},
  year = 2009
}

@article{estrada,
  title={Communicability in complex networks},
  author={Estrada, Ernesto and Hatano, Naomichi},
  journal={Physical Review E},
  volume={77},
  number={3},
  doi = {10.1103/PhysRevE.77.036111},
  pages={036111},
  year={2008},
  publisher={APS}
}

@article{freeman,
  title={A set of measures of centrality based on betweenness},
  author={Freeman, Linton C},
  journal={Sociometry},
  pages={35--41},
  year={1977},
  doi={10.2307/3033543},
  publisher={JSTOR}
}

@article{ruhnau2000,
  title={Eigenvector-centrality—a node-centrality?},
  author={Ruhnau, Britta},
  journal={Social networks},
  volume={22},
  number={4},
  url={10.1016/S0378-8733(00)00031-9},
  pages={357--365},
  year={2000},
  publisher={Elsevier}
}

@article{chen2018,
  title={Complex network comparison based on communicability sequence entropy},
  author={Chen, Dan and Shi, Dan-Dan and Qin, Mi and Xu, Si-Meng and Pan, Gui-Jun},
  journal={Physical Review E},
  volume={98},
  number={1},
  doi = {10.1103/PhysRevE.98.012319},
  pages={012319},
  year={2018},
  publisher={APS}
}

@article{Gfeller2007,
  title = {Spectral Coarse Graining of Complex Networks},
  author = {Gfeller, David and De Los Rios, Paolo},
  journal = {Phys. Rev. Lett.},
  volume = {99},
  issue = {3},
  pages = {038701},
  numpages = {4},
  year = {2007},
  month = {Jul},
  publisher = {American Physical Society},
  doi = {10.1103/PhysRevLett.99.038701},
  url = {https://link.aps.org/doi/10.1103/PhysRevLett.99.038701}
}

@article{Krapivsky2001,
  title = {Degree Distributions of Growing Networks},
  author = {Krapivsky, P. L. and Rodgers, G. J. and Redner, S.},
  journal = {Phys. Rev. Lett.},
  volume = {86},
  issue = {23},
  pages = {5401--5404},
  numpages = {0},
  year = {2001},
  month = {Jun},
  publisher = {American Physical Society},
  doi = {10.1103/PhysRevLett.86.5401},
  url = {https://link.aps.org/doi/10.1103/PhysRevLett.86.5401}
}

@article{gfeller2008spectral,
  title={Spectral coarse graining and synchronization in oscillator networks},
  author={Gfeller, David and De Los Rios, Paolo},
  journal={Physical review letters},
  volume={100},
  number={17},
  pages={174104},
  year={2008},
  publisher={APS}
}

@article{hoel2020evolution,
  title={Evolution leads to emergence: An analysis of protein interactomes across the tree of life},
  author={Hoel, Erik and Klein, Brennan and Swain, Anshuman and Griebenow, Ross and Levin, Michael},
  journal={bioRxiv},
  year={2020},
  publisher={Cold Spring Harbor Laboratory}
}

@article{kaiser2004spatial,
  title={Spatial growth of real-world networks},
  author={Kaiser, Marcus and Hilgetag, Claus C},
  journal={Physical Review E},
  volume={69},
  number={3},
  pages={036103},
  year={2004},
  publisher={APS}
}

@article{shang2017fitness,
  title={Fitness networks for real world systems via modified preferential attachment},
  author={Shang, Ke-ke and Small, Michael and Yan, Wei-sheng},
  journal={Physica A: Statistical Mechanics and its Applications},
  volume={474},
  pages={49--60},
  year={2017},
  publisher={Elsevier}
}

@article{barabasi1999emergence,
  title={Emergence of scaling in random networks},
  author={Barab{\'a}si, Albert-L{\'a}szl{\'o} and Albert, R{\'e}ka},
  journal={science},
  volume={286},
  number={5439},
  pages={509--512},
  year={1999},
  publisher={American Association for the Advancement of Science}
}

@article{koschutzki2008centrality,
  title={Centrality analysis methods for biological networks and their application to gene regulatory networks},
  author={Kosch{\"u}tzki, Dirk and Schreiber, Falk},
  journal={Gene regulation and systems biology},
  volume={2},
  pages={GRSB--S702},
  year={2008},
  publisher={SAGE Publications Sage UK: London, England}
}

@article{leydesdorff2007betweenness,
  title={Betweenness centrality as an indicator of the interdisciplinarity of scientific journals},
  author={Leydesdorff, Loet},
  journal={Journal of the American Society for Information Science and Technology},
  volume={58},
  number={9},
  pages={1303--1319},
  year={2007},
  publisher={Wiley Online Library}
}

@inproceedings{zhang2014finding,
  title={Finding most vital node in satellite communication network},
  author={Zhang, Xiao Juan and Wang, Zu Lin and Zhang, Zhi Xia},
  booktitle={Applied Mechanics and Materials},
  volume={635},
  pages={1136--1139},
  year={2014},
  organization={Trans Tech Publ}
}

@article{onnela2007structure,
  title={Structure and tie strengths in mobile communication networks},
  author={Onnela, J-P and Saram{\"a}ki, Jari and Hyv{\"o}nen, Jorkki and Szab{\'o}, Gy{\"o}rgy and Lazer, David and Kaski, Kimmo and Kert{\'e}sz, J{\'a}nos and Barab{\'a}si, A-L},
  journal={Proceedings of the national academy of sciences},
  volume={104},
  number={18},
  pages={7332--7336},
  year={2007},
  publisher={National Acad Sciences}
}

\end{document}